# Removal of persistent organic contaminants from wastewater using a hybrid electrochemical-granular activated carbon (GAC) system

Giannis-Florjan Norra [a,b], Jelena Radjenovic [a,c*]

[a]Catalan Institute for Water Research (ICRA), Emili Grahit 101, 17003 Girona, Spain

[b]University of Girona, Girona, Spain

[c]Catalan Institution for Research and Advanced Studies (ICREA), Passeig Lluís Companys 23, 08010 Barcelona, Spain

* Corresponding author:

Jelena Radjenovic,

Catalan Institute for Water Research (ICRA), Emili Grahit 101, 17003 Girona, Spain

Phone: + 34 972 18 33 80; Fax: +34 972 18 32 48; E-mail: jradjenovic@icra.cat

**Abstract**

A three-dimensional (3D) electrochemical flow-through reactor equipped with GAC packed bed, polarized by the electric field, was evaluated for the removal of persistent organic contaminants from real sewage effluent. The performance of the reactor was investigated for 27 consecutive runs at two anodic current densities, i.e., low current density (LCD) of 15 A m$^{-2}$, and high current density (HCD) of 100 A m$^{-2}$. In the HCD experiments, the adsorption ability of saturated GAC was increased, mainly due to the increase in the mesoporosity of GAC. A synergy between electrosorption/adsorption on GAC and electrooxidation was observed in terms of the removal of all target pollutants. DEET presented the highest synergy, ranging from 40-57%, followed by iopromide (22-46%), carbamazepine (15-34%) and diatrizoate (4-30%). The addition of GAC decreased the concentrations of toxic chlorate and perchlorate by 2-fold and 10-fold, respectively, due to their electrosorption on GAC. Also, 3D electrochemical system yielded lower concentrations of adsorbable organic iodide (AOI) and adsorbable organic chlorine (AOCl). Thus, addition of low amounts of GAC in electrochemical systems may be a low-cost and simple way of minimizing the formation and final effluent concentrations of toxic halogenated byproducts.

# Introduction

Electrochemical processes are among the most promising and emerging technologies in the field of (waste)water treatment, presenting various advantages, such as no addition of chemicals, versatility and robustness of operation when treating contaminated water of different origin, and compact design, which makes them very well suited for decentralized and distributed treatment systems (Radjenovic and Sedlak, 2015). Nevertheless, the main drawback of conventional, two-dimensional (2D) electrochemical reactors are high mass transfer limitations governing the oxidation or reduction of the contaminants at the electrode surface, resulting in high energy consumption and operational costs (Cañizares et al., 2004; Zhang et al., 2013). In addition, anodic oxidation of chloride that is present in virtually any water source and wastewater leads to the formation of chlorine (i.e., $Cl_2$, $HOCl/OCl^-$), a long-lived oxidant species that reacts with the organic matter to form persistent and toxic chlorinated byproducts (Bagastyo et al., 2011; Radjenovic and Sedlak, 2015). The removal of organochlorine byproducts, both chlorinated aromatics and low molecular-weight compounds (e.g., trihalomethanes, THMs) is typically achieved using granular activated carbon (GAC) (Cuthbertson et al., 2019; Jiang et al., 2017). Bench-scale studies demonstrated a successful reduction of chlorinated byproducts formed in electrochemical treatment by a GAC post-treatment (Rajkumar et al., 2005), as well as overall enhanced performance of electrochemical treatment by adding GAC as both pre- and post-treatment (Rogers et al., 2018).

Due to its good electrical conductivity and high specific surface area, GAC was also employed as a particle electrode to construct three-dimensional (3D) electrochemical systems, resulting in a higher electroactive surface area and shorter distances for mass transfer of trace contaminants (Zhang et al., 2013). Great majority of studies on hybrid GAC-electrochemical systems was based on GAC bed placed between the vertically positioned anode and cathode and in direct contact with one or both terminal electrodes, typically cathode (Kılıç et al., 2007; Zhan et al., 2019). However, direct polarization of GAC at higher currents and potentials can cause its attrition and compromise the treatment performance (Gedam and Neti, 2014; Kılıç et al., 2007). If the applied electric field is large enough, conductive GAC particles do not need to be in contact with the terminal electrodes and are polarized due to the shift in their electric charge, converting each GAC particle into a charged microelectrode, with one anodic and one cathodic side (Zhang et al., 2013). Charged GAC particles can thus induce electrosorption/desorption of contaminants and their Faradaic reactions, thus enabling electrolytic degradation of the adsorbed organics (McQuillan et al., 2018). In addition, GAC can enhance the system performance and increase the production of hydroxyl radicals (HO$^\bullet$) through the formation of $H_2O_2$ at cathodically polarized particles (Bañuelos et al., 2014) and its decomposition by GAC to HO$^\bullet$ (Fortuny et al., 1998). The amount of $H_2O_2$ formed at the activated carbon cathode was found to increase linearly with the increase in dissolved oxygen (DO) content (Bañuelos et al., 2014). The anode-cathode flow direction is often the preferred configuration for 3D flow-through reactors as it minimizes the impact of undesirable side reaction (e.g., $H_2$ evolution) and reduces the risk of re-oxidation of the already reduced products (e.g., $NO_2^-$, NO) (Garcia-Segura et al., 2020). Previously, addition of GAC packed bed decreased the energy consumption of electrochemical greywater treatment from 46 kWh $kg_{COD}^{-1}$ in a conventional 2D system to 30 kWh $kg_{COD}^{-1}$ removed in the 3D electrochemical system (Andrés García et al., 2018).

In this study, we have investigated for the first time the performance of the flow-through 3D electrochemical reactor equipped with GAC packed bed polarized by the electric field, in removing persistent organic contaminants from real sewage effluent. To promote the *in-situ* formation of $H_2O_2$, we used a flow-through reactor with boron-doped diamond (BDD) mesh anode and stainless-steel mesh cathode orientated in anode-cathode flow direction, and influent feed from the bottom of the reactor, thus enabling a continuous supply of DO to the GAC packed bed. As model contaminants, we selected compounds known to be persistent to oxidative degradation, such as iodinated contrast media diatrizoate and iopromide, anti-epileptic drug carbamazepine and insecticide N,N-diethyl-meta-toluamide (DEET) (Dickenson et al., 2009). The experiments were conducted using saturated GAC over multiple subsequent cycles and at low and high applied currents, to study its impact on the process performance and GAC surface characteristics. To gain insight into the potential of the GAC packed bed to minimize the presence of chlorinated byproducts in the system, we analyzed adsorbable organic halogen (AOX) in both 2D and 3D electrochemical systems. We also evaluated the formation of inorganic chlorinated byproducts chlorate ($ClO_3^-$) and perchlorate ($ClO_4^-$). This study provides new insights into the fate of organic pollutants in a hybrid electrochemical-GAC system when working with real wastewater, impact of the applied current on their removal, GAC surface characteristics and presence of organic and inorganic chlorinated byproducts.

## Materials and methods

### Chemicals

Analytical standards for diatrizoate (DTR), carbamazepine (CBZ), DEET and iopromide (IPM) were purchased from Sigma-Aldrich. The physico-chemical properties of target contaminants are summarized in **Table S1**. All solutions were prepared using analytical grade reagents and Milli-Q water. Secondary treated sewage effluent was collected from local municipal wastewater treatment plant (WWTP) in Girona, Spain. Characteristics of the sewage effluent are summarized in **Table S2**.

**Experimental Setup**

Experiments were conducted in a flow-through electrochemical reactor equipped with a mesh Nb/BDD anode ($80\times80\times1.3$ mm, DIACHEM® electrode, Condias, Germany), both sides of the niobium substrate coated with 4 µm of BDD thin film, and a stainless-steel cathode of the same dimensions. The GAC used was CG 1000, provided by ChiemiVall, S.L. (Spain) with granule diameter 2.4—4.8 mm and apparent density of $500\pm30$ kg m$^{-3}$. Prior to the experiments, GAC was saturated using sewage effluent until its adsorption capacity for the removal of chemical oxygen demand (COD) was reduced to $\leq15\%$ of COD removal obtained. In the case of 3D electrochemical oxidation (3D ELOX) reactor, a packed bed of GAC was placed in between anode and cathode. The total volume of 2D electrochemical system was 250 mL, whereas 3D packed bed comprised approximately 65 mL, i.e., 26% of the total volume. BDD anode was positioned at the bottom of the reactor, and wastewater was circulated in the direction from anode to cathode. The experiments were conducted in batch mode using 2 L of secondary sewage effluent amended with the target contaminants at 2 µM initial concentration, and with the recirculation flowrate of 475 mL min$^{-1}$. The duration of each experiment was 6 h. Chronopotentiometric experiments were conducted at 96 and 640 mA of applied anodic current (i.e., current densities calculated versus the projected anode surface area of 15 and 100 A m$^{-2}$, respectively) using a BioLogic multi-channel potentiostat/galvanostat VMP-300 and a leak-free 1 mm diameter Ag/AgCl reference electrode (Harvard Apparatus) placed in the proximity of the anode. A scheme of the experimental set-up used is illustrated in **Figure S1**. All potentials in this manuscript are expressed versus Standard Hydrogen Electrode (V/SHE). To examine the synergy between adsorption on GAC and electrooxidation/electrosorption, experiments were performed in the open circuit (OC) mode, i.e., without the application of current (ADS), and in an electrochemical system without the addition of the GAC bed (2D ELOX). All samples were quenched with methanol and frozen immediately after sampling. The performance of 3D ELOX system was evaluated in 27 consecutive runs applying low current density (LCD) of 15 A m$^{-2}$ and high current density (HCD) of 100 A m$^{-2}$, to evaluate the impact of anodic current density and electric field strength on the process performance. Runs 12 and 28 were conducted in the open circuit to determine the evolution of the adsorption capacity and regeneration efficiency of GAC. To prevent the growth of the biomass at the surface of the GAC particles, secondary effluent was previously autoclaved, whereas GAC was autoclaved once per week (121°C, 10 min). All experiments were performed in duplicate and the results are expressed as mean values with their standard deviations.

**Analysis**

Dissolved organic carbon (DOC) was determined as non-purgeable organic carbon (NPOC) using a total organic carbon analyzer (TOC-V CSH, Shimadzu, Kyoto, Japan). COD was measured using COD cuvette test kits (15-150 mg/L O$_2$LCK 314) provided by Hach using a spectrophotometric method (APHA, 2017). Target organic contaminants were analyzed in selected reaction monitoring (SRM) mode, using a 5500 QTRAP hybrid triple quadrupole-linear ion trap mass spectrometer (Applied Biosystems, Foster City, CA, USA) with a turbo Ion Spray source, coupled to a Waters Acquity Ultra-Performance$^{TM}$ liquid chromatograph (UPLC) (Milford, MA, USA). Details of the analytical method employed are summarized in **Text S1** and **Table S3**.

Free available chlorine (FAC) and total chlorine (sum of FAC and combined chlorine) were measured with the N,N-diethyl-p-phenylenediamine (DPD) ferrous titrimetric method (APHA, 2017) using LCK 310 cuvette tests provided by Hach Lange Spain Sl (Barcelona), immediately after sampling. The concentrations of chloride, chlorate and perchlorate were determined using high-pressure ion chromatography (HPIC) system (Dionex ICS-5000). The Langmuir surface area and pore size distribution were determined by $N_2$ adsorption-desorption at 77K using Micromeritics ASAP® 2420 Accelerated Surface Area and Porosimetry System.

Halogen-specific AOX analyses, i.e. adsorbable organic chlorine (AOCl), adsorbable organic bromine (AOBr) and adsorbable organic iodine (AOI) were conducted according to the previously published methodology (von Abercron et al., 2019). 20 mL samples were quenched with 40 μL of 40 g L$^{-1}$ $Na_2SO_3$ per 3 mg L$^{-1}$ of free chlorine, acidified to pH 2 and extracted on two activated carbon cartridges using a TXA-04 AOX Adsorption Unit. Afterwards, the cartridges were washed with 10 mL of 10 mM $NaNO_3$ solution to remove the inorganic halides. The cartridges were then combusted at 1000ºC in a horizontal furnace AQF-2100H equipped with GA-210 Gas adsorption unit (COSA-Mitsubishi). The halides formed were analyzed using a Dionex Integrion HPIC system. To analyze the surface of GAC granules, scanning electron microscopy (SEM) analyses were performed on a FEI Quanta FEG (pressure: 70Pa; HV: 20kV; and spot: four).

## Results and Discussion

The apparent removal rate constants of the contaminant, COD and DOC removal in 2D ELOX ($k_{2D}$), 3D ELOX ($k_{3D}$) and ADS system ($k_{ADS}$) were expressed in $h^{-1}$ and estimated using the linear regression first-order decay model fit to the data, with coefficient $R^2 \geq 0.9$. For a limited number of experiments in the ADS system, first-order rate constants were calculated using the initial stages of adsorption (i.e., first 2 h) instead of the data obtained during the entire 6 h of the experiments, to allow the comparison between different experiments. Good fits obtained for the removal of target contaminants on the saturated GAC in flow-through mode suggest that their adsorption was controlled by the diffusion at the boundary layer, similar to the previously reported data for pilot- and full-scale GAC (Golovko et al., 2020). The synergy between adsorption (ADS) and electrochemical oxidation (2D ELOX) in the 3D system (3D ELOX) can be calculated using the following equation:

$$Synergy\ (\%) = \frac{k_{3D} - k_{2D} - k_{ADS}}{k_{3D}} \qquad \text{(eq. 1)}$$

### *Impact of electric field on the structural properties of GAC*

To study the impact of electrochemical polarization of GAC on its surface properties, specific surface area and porosity of the initial (i.e., saturated GAC) and GAC used in the LCD and HCD experiments were investigated. In the LCD experiment, Langmuir surface area of the granules decreased from 1129 to 998 $m^2\,g^{-1}$ (**Table 1**). This decrease was more pronounced in the HCD experiment, with the Langmuir surface area of the GAC after 27 runs lowered to 570 $m^2 g^{-1}$. The T-plot Micropore area, which corresponds to the part of Langmuir Area that is microporous, decreased from 1108 $m^2 g^{-1}$ for the initial GAC to 978 $m^2 g^{-1}$ and 557 $m^2 g^{-1}$ after the LCD and HCD experiments, respectively, indicating a significantly larger decrease in microporosity of GAC when higher currents are applied. The determined Barrett-Joyner-Halenda (BJH) average pore diameters were similar for the saturated GAC (36.5 Å), and GAC after the LCD (35.8 Å) and the HCD experiments (36.3 Å). The density functional theory (DFT) pore size analysis demonstrated that there was an increase in the area of larger pores (i.e., $\geq$22 Å), which was again more pronounced in the HCD experiment (i.e., from 3 to 11 $m^2 g^{-1}$) compared with the LCD experiment (from 3 to 6 $m^2 g^{-1}$). Previous studies indicated a higher accuracy and reliability of the DFT method compared with the BJH method (Bardestani et al., 2019). Thus, electrochemical polarization of GAC in the HCD experiments led to a more pronounced increase in the average GAC pore size as determined by the DFT method, lower T-plot micropore area and lower Langmuir surface area in the. Larger GAC pore size after the HCD experiment was also confirmed by the SEM images (**Figure S2**).. Although there was no visible attrition of GAC during the LCD and HCD experiments it is likely that at the higher applied current led to GAC attrition, even under bipolar polarization. Furthermore, the measured pore widening and a pronounced decrease in Langmuir surface area in the HCD experiments evidence changes in GAC porous structure, likely due to the formation of $H_2O_2$ and reactive oxygen species at the GAC surface. Larger pores generated under electrochemical polarization of GAC may be beneficial for the removal of organic pollutants. Yet, further studies should be conducted to evaluate whether bipolar polarization of GAC over long-term leads to mass losses due to GAC degradation.

**Table 1.** Structural properties of the initial (i.e., saturated) GAC and GAC used in 3D ELOX reactor in LCD and HCD experiments.

|  | Saturated GAC | LCD GAC | HCD GAC |
|---|---|---|---|
| **Langmuir Surface Area (m²/g):** | 1129 | 998 | 570 |
| **The T-Plot Micropore Area (m²/g)** | 1108 | 978 | 557 |
| **DFT Total Area in Pores >22Å (m²/g)** | 3 | 6 | 11 |
| **BJH Adsorption average pore diameter (4V/A) (Å):** | 36.5 | 35.8 | 36.3 |

### *Removal of trace organic contaminants in low and high current density experiments*

**Table S4** summarizes the $k_{2D}$, $k_{3D}$ and $k_{ADS}$ values for target contaminants in the LCD and HCD experiments. Among the target contaminants, carbamazepine and DEET have the highest octanol-water distribution coefficient (logD), 2.77 and 2.50 (**Table S1**) and are thus expected to be adsorbed to the highest degree. The $k_{ADS}$ of DEET was 0.19 $h^{-1}$, similar to the observed $k_{ADS}$ for carbamazepine (i.e., 0.33 $h^{-1}$), significantly higher than in the case of iodinated contrast media (ICM) (i.e., 0.08 $h^{-1}$ for iopromide and 0.02 $h^{-1}$ for diatrizoate) (**Table S4**). Furthermore, adsorption of larger molecules of iopromide (molecular weight, MW of 791.1 g $mol^{-1}$) and diatrizoate (613.9 g $mol^{-1}$) on the micropores of GAC is subject to more pronounced diffusion limitations that in the case of smaller molecules of carbamazepine (236.3 g $mol^{-1}$) and DEET (191.3 g $mol^{-1}$) (Li et al., 2018).

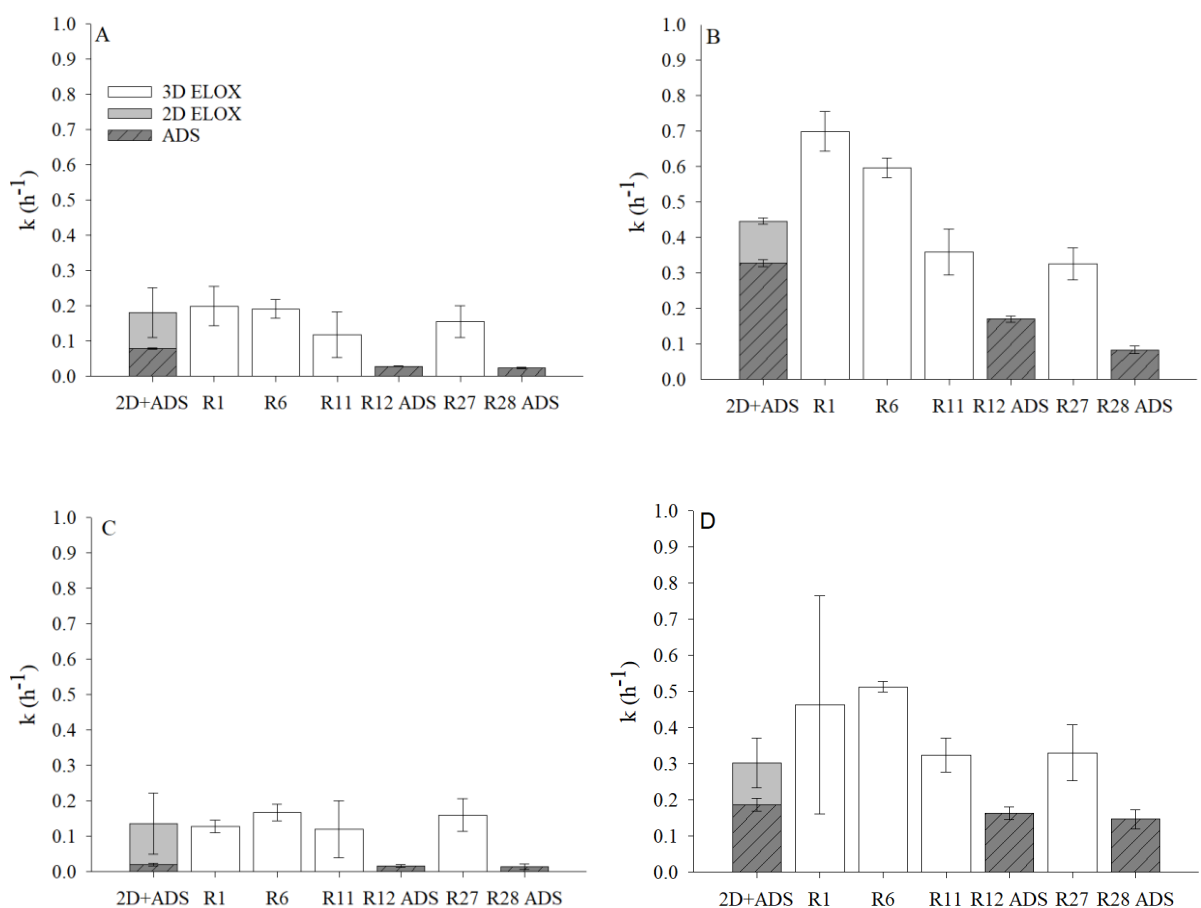

**Figure 1**. Apparent first-order removal rate constants (k, h$^{-1}$) for: **A)** IPM, **B)** CBZ, **C)** DTR, and **D)** DEET in the LCD experiments (15 A m$^{-2}$).

**Figure 1** shows the removal rate constants observed in the LCD experiments. For carbamazepine and DEET, synergy in the 3D ELOX system was observed in the first runs with saturated GAC (i.e., run 1 and 6), but it was diminished in the subsequent runs, resulting in even negative synergy (**Figure 1**). Although higher rate constants were obtained for diatrizoate in runs 6 and 27 compared with the sum of 2D ELOX and ADS removal rate constants (i.e., $k_{2D}+k_{ADS}=0.12$ h$^{-1}$+0.02h$^{-1}$), there was no steady trend in $k_{3D}$ that would evidence the synergy between electrochemical oxidation and electrosorption. The removal rates of the other target pollutants in the LCD experiment were significantly decreased in the subsequent application cycles, i.e., from 0.20 h$^{-1}$ (run 1) to 0.16 h$^{-1}$ (run 27) for iopromide, 0.70 h$^{-1}$ (run 1) to 0.33 h$^{-1}$ (run 27) for carbamazepine, and 0.46 h$^{-1}$ (run 1) to 0.33 h$^{-1}$ (run 27) for DEET. The anode potential in these experiments was 2.1 V/SHE, which is lower than the thermodynamic standard potential for HO$^{\bullet}$ formation, i.e., 2.38 V/SHE (Kapałka et al., 2009). Furthermore, the potential difference over the microconductor, $\Delta U_{microconductor}$, in the LCD experiment, of 0.9-1.8 V was likely insufficient to induce efficient charge separation in GAC particles. $\Delta U_{microconductor}$ depends on the electric field gradient between the terminal electrodes and the length of the conductor. If the applied potential is high enough, the conductive particles in the electric field become polarized, without being in contact with any of the main electrodes. For Faradaic redox reactions to occur at the surface of the microelectrodes, the potential difference over the microconductor needs to be high enough to reach at least the value of the redox potential of the considered redox reaction, e.g., cathodic reduction of oxygen to H$_2$O$_2$ (McQuillan et al., 2018; Rahner et al., 2002). The saturation of GAC was also evident from the gradual increase in ohmic resistance in the 3D ELOX system, with total cell potential increasing from 3.6 V in run 1 to 6.9 V in run 27. Given that the conductivity of wastewater was relatively constant in the LCD experiment, this increase in total cell potential could be assigned to a continuous increase in resistance of GAC packed bed.

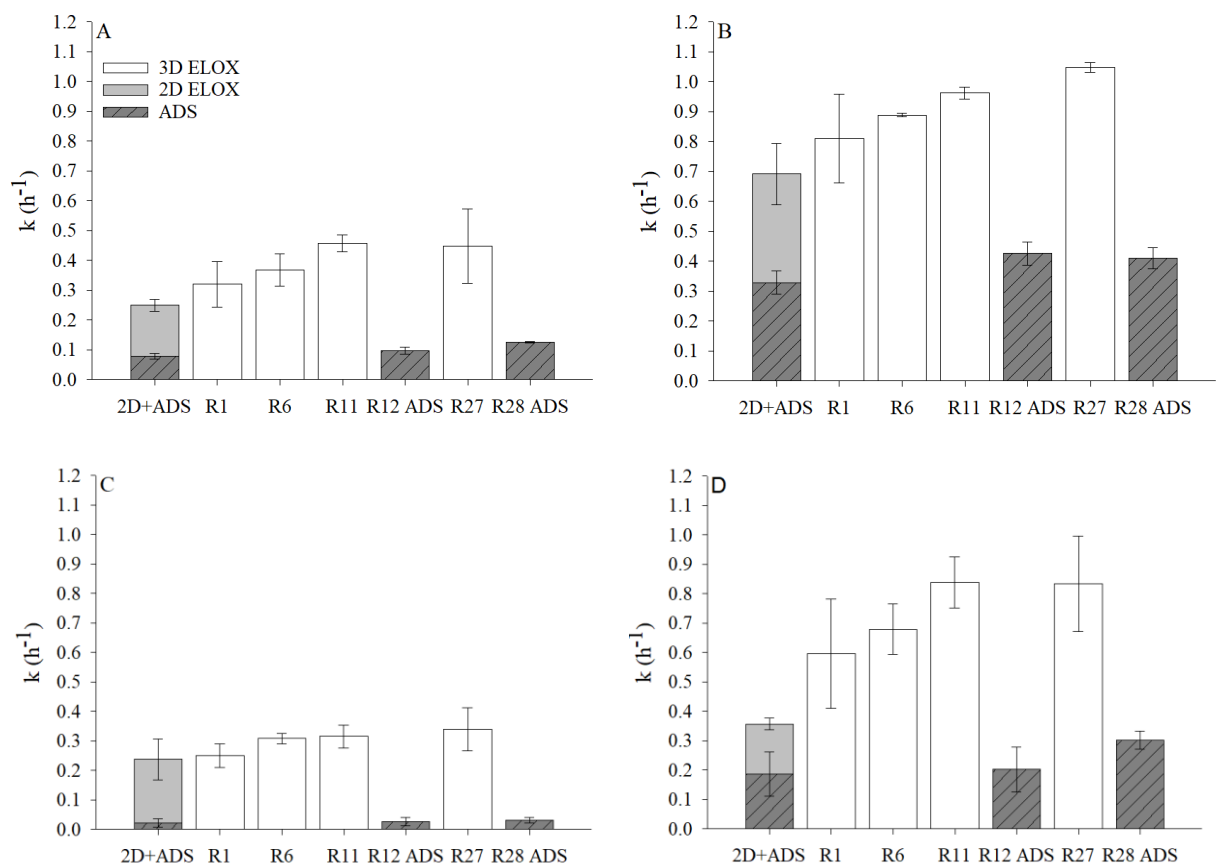

**Figure 2**. Apparent first-order removal rate constants (k, h⁻¹) for: **A)** IPM, **B)** CBZ, **C)** DTR, and **D)** DEET in the HCD experiments (100 A m⁻²).

In the HCD experiments, synergy between electrochemical oxidation and adsorption/electrosorption was observed for all contaminants, with a gradual increase in their removal rate constants and observed synergy values with the conducted runs (**Figure 2**). Synergy of electrosorption and electrooxidation in 3D ELOX system was 22% for iopromide removal in run 1 and was steadily increased to 44-46% (run 11 and run 27). Similar results were obtained for carbamazepine (28-34% synergy), diatrizoate 24-30% synergy) and DEET (57.2-57.4% synergy) in runs 11-27 (**Table S5**). In the pH range of the experiment (pH 7.2.-9.1), iopromide, carbamazepine and DEET are neutral molecules whereas diatrizoate is negatively charged. Electrosorption of uncharged species is still likely to occur due to their polarizability by the external electric field and formation of dipoles (Niu and Conway, 2002; Thamilselvan et al., 2018). The highest molecular polarizability is predicted for iopromide (i.e., 56.4 $Å^3$), followed by diatrizoate (39.8 $Å^3$), carbamazepine (28 $Å^3$) and DEET 23.3 $Å^3$ (**Table S1**). Thus, high synergy values of iopromide may be a consequence of its enhanced electrosorption on the polarized GAC particles. However, the highest increase in $k_{3D}$ values during repeated applications was noted for DEET, for which the removal rate constant was significantly increased already in run 1, from $k_{2D}+k_{ADS}=017+0.19\,h^{-1}$ to $k_{3D}=0.60\,h^{-1}$, resulting in 40% of synergy. This could be explained by efficient adsorption of a relatively hydrophobic (log D=2.50) and small molecule of DEET (191.3 g $mol^{-1}$) on GAC and its enhanced degradation by the $OH^{•}$, generated in HCD experiment ($k_{OH}=5.00\times10^9\,M^{-1}\,s^{-1}$, **Table S6**).

In the HCD experiment, BDD anode potential was 2.68 V/SHE, thus ensuring the electrogeneration of $OH^{•}$ (Kapałka et al., 2010). Yet, the concentration of anodically formed $OH^{•}$, is expected to decrease exponentially with the distance from the electrode surface and reach zero already at 20 nm distance (Groenen-Serrano et al., 2013; Kapałka et al., 2009). Thus, they are not expected to contribute to the synergy observed in the 3D ELOX and GAC regeneration. The potential difference over the microconductor, $\Delta U_{microconductor}$ in the HCD experiments ranged from 3.6 to 6.9 V, significantly higher than in the LCD experiments, and it ensured electrocatalytic activity of GAC packed bed (Andrés García et al., 2018; Pedersen et al., 2019; Polcaro et al., 2000; Zhu et al., 2011). Intensive generation of oxygen at the BDD anode placed below the GAC packed bed likely enhanced the production of $H_2O_2$ at the cathodically polarized sides of GAC particles, according to the eq. 2:

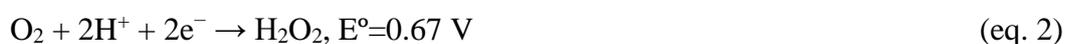

$$O_2 + 2H^+ + 2e^- \rightarrow H_2O_2,\ E°=0.67\ V \qquad\qquad\qquad (eq.\ 2)$$

Previous studies have demonstrated enhanced electrogeneration of $H_2O_2$ in the anode-cathode configuration due to the reduction of anodically generated $O_2$ at the cathode (Zhou et al., 2018). Regardless of the application of current, the formed $H_2O_2$ can be decomposed by GAC particles to $OH^\bullet$ due to the presence of polyaromatic moieties and functional groups in GAC (Zhang et al., 2013; Zhu et al., 2011). The decomposition of $H_2O_2$ to $OH^\bullet$ on activated carbon was first reported by Kimura and Miyamoto (Kimura and Miyamoto, 1994), and the postulated mechanism includes the dissociation of the $H_2O_2$ to peroxyl radical ($OOH^\bullet$) at the carbon surface, which forms superoxide radical ion (eq. 3):

$$OOH^\bullet \rightleftharpoons H^+ + O_2^{\bullet-} \qquad\qquad\qquad (eq.\ 3)$$

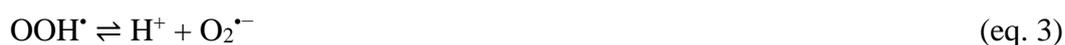

The superoxide radical ion reacts further with $H_2O_2$ to form $OH^\bullet$ (eq. 4) (Kimura and Miyamoto, 1994):

$$H_2O_2 + O_2^{\bullet-} \rightarrow OH^- + OH^\bullet + O_2 \qquad\qquad\qquad (eq.\ 4)$$

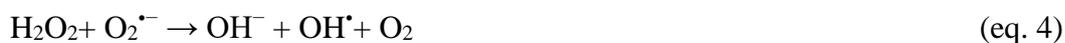

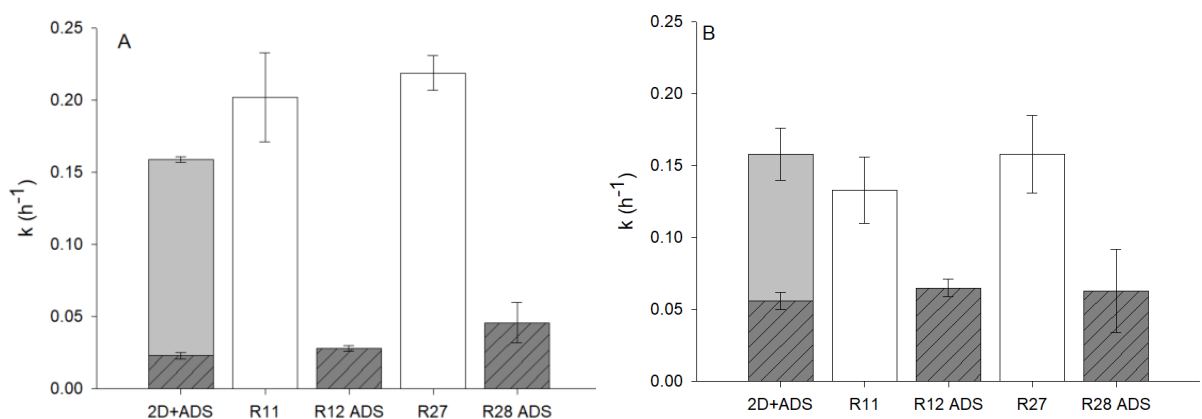

**Figure 3**. Apparent first-order removal rate constants (k, $h^{-1}$) for: **A)** COD, and **B)** TOC removal in the HCD experiment.

The generation of $H_2O_2$ and its activation to $OH^\bullet$ are dependent on the carbon porosity, chemical properties of the surface, pH and other factors (Anfruns et al., 2014; Bach and Semiat, 2011). Thus, polarization of GAC can further contribute to the degradation of the adsorbed pollutants and organic matter, and regeneration of GAC in the 3D ELOX system. Higher synergy of electrosorption and electrochemical degradation in the HCD experiment was also evident from the measured COD removal rates (**Figure 3A**). On the other hand, TOC removal remained relatively constant and no synergy could be observed in the HCD experiments (**Figure 3B**). This could be a consequence of the formation of halogenated organic byproducts (i.e., AOX) that are more resistant to oxidative degradation (Lan et al., 2017).

In the HCD experiments, only in the case of carbamazepine, a removal higher than 90% was achieved. The electric energy per order ($E_{EO}$), i.e., energy required to reduce the concentration of a contaminant by one order of magnitude in a unit volume of treated solution) (Andrés García et al., 2018), ranged from 14.3-23.1 kWh m$^{-3}$ for the removal of carbamazepine in 3D ELOX experiments, while it was 42.3 kWh m$^{-3}$ in the 2D ELOX process (**Table S7**). High $E_{EO}$ values determined are a consequence of low electric conductivity of the secondary effluent (1.1 mS cm$^{-1}$), which resulted in very high total cell potential in both 2D and 3D ELOX systems (15-25 V). For the rest of the target contaminants, given their high persistency to oxidative degradation, lower removals were achieved and thus $E_{Eo}$ was not calculated. It should be noted here that high values for $E_{EO}$ were a consequence of working with low conductivity wastewater (1.1 mS cm$^{-1}$), and significant spacing between the GAC packed bed and the anode (1 cm) required to accommodate the Ag/AgCl reference electrode. The ohmic resistance of the system can be significantly reduced by reducing the interelectrode spacing. Furthermore, the electric resistance of the employed GAC can have a significant impact not only on the ohmic drop of the reactor but also on the performance of electrochemical regeneration, as observed previously (Narbaitz and Karimi-Jashni, 2009).

In both LCD and HCD experiments, runs 12 and 28 were conducted in the OC, i.e., without the application of current, to investigate the impact of electrochemical polarization of GAC on its adsorption performance. In the LCD treatment, most of the rate constants for the runs were similar or slightly decreased from the initial saturated GAC to run 12 and run 28 (**Figure 1, Table S4**). For example, $k_{ADS}$ of iopromide were decreased from 0.08 $h^{-1}$ to 0.03 $h^{-1}$ (run 12) and 0.023 $h^{-1}$ (run 28) when GAC was used in the LCD experiment. On the other hand, in the HCD treatment, $k_{ADS}$ was increased from the adsorption on the initial GAC to adsorption on used GAC runs 12 and 28 (**Figure 2, Table S4**). More notable differences were observed for DEET, for which the $k_{ADS}$ remained constant in the LCD experiment over subsequent runs, whereas in the HCD experiment it was increased from 0.19 $h^{-1}$ to 0.30 $h^{-1}$ in run 28. Also, $k_{ADS}$ of carbamazepine was increased from 0.33 $h^{-1}$ to 0.43 $h^{-1}$ (run 28). Increased amount of mesopores due to the electrochemical polarization of GAC at higher currents would mean better accessibility of the contaminants and easier regeneration compared with the micropores (Xiao and Hill, 2019). Also, pore blockage and competitive adsorption of organic matter with the adsorbed contaminants was reported to be less pronounced in activated carbons with larger pore sizes (Ding et al., 2008; Ebie et al., 2001; F. Li et al., 2003; Li et al., 2018; Q. Li et al., 2003; Pelekani and Snoeyink, 1999). Considering the above, the wider pores obtained in the HCD experiment should enhance the adsorption of the trace organic contaminant and increase the synergies observed in the 3D ELOX system. Previously, higher adsorption rates of rhodamine-B and ibuprofen were observed for activated carbons with broadened hierarchical pore distribution (Li et al., 2018; Mestre et al., 2009). Furthermore, mesopores are important as transport arteries through which the adsorbate molecules reach the micropores (Moreno-Castilla and Rivera-Utrilla, 2001). The adsorption of the different contaminants depends on their molecular weight in correlation with the pore size distribution of the activated carbon (Li et al., 2018). The micropososity should be large enough to ensure accessibility of big MW compounds (Mestre et al., 2009).

*Formation of organic and inorganic chlorinated byproducts*

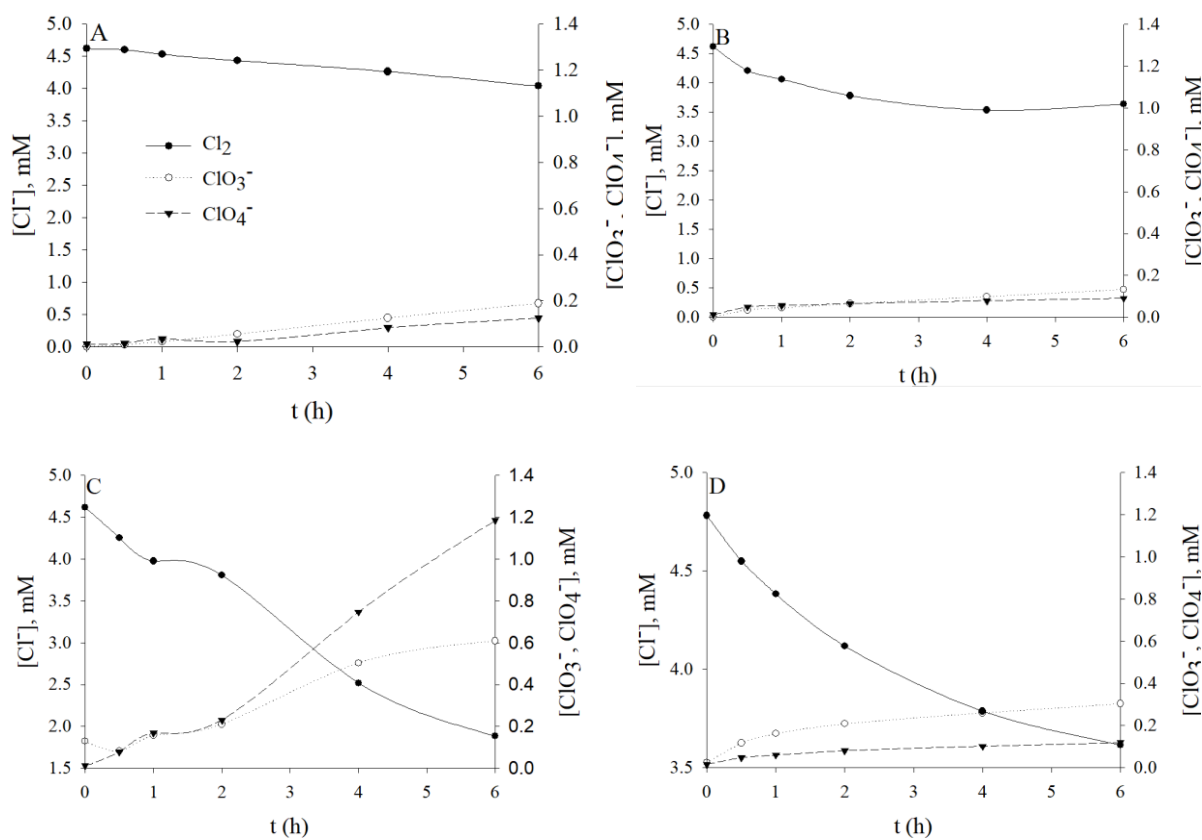

**Figure 4.** Measured concentrations of Cl$^-$, ClO$_3^-$, and ClO$_4^-$ in: **A)** 2D ELOX LCD, **B)** 3D ELOX LCD, **C)** 2D ELOX HCD, and **D)** 3D ELOX HCD experiments.

In the presence of chloride, anodic polarization of BDD leads to the formation of free chlorine (HOCl/OCl$^-$), chlorine radicals and in the case of highly saline solutions, gaseous ClO$_2$ and Cl$_2$O (Mostafa et al., 2018). Faster decrease in Cl$^-$ was observed in the HCD experiments (0.07 h$^{-1}$) compared with LCD experiments (0.02 h$^{-1}$) in the 2D ELOX system in (**Figure 4**). The presence of GAC packed bed in 3D ELOX systems did not have a significant impact on the kinetics of chloride oxidation (0.05 h$^{-1}$ and 0.08 h$^{-1}$ in LCD and HCD experiments, respectively). Yet, active chlorine was present only in the 2D ELOX systems, where residual chlorine was measured in concentrations of up to 6.04 and 7.11 mg L$^{-1}$ in LCD and HCD experiments, respectively. In the case of 3D ELOX, residual free chlorine (Cl$_2$/HOCl/OCl$^-$) was measured only in traces (i.e., 0.53-1.07 mg L$^{-1}$), as it can be seen in **Figure 5**. It should be noted here that due to the high reactivity of active chlorine with the organic matter, the measurements of free chlorine using the DPD method represent residual Cl$_2$/HOCl/OCl$^-$, in the system, and not the amount generated.

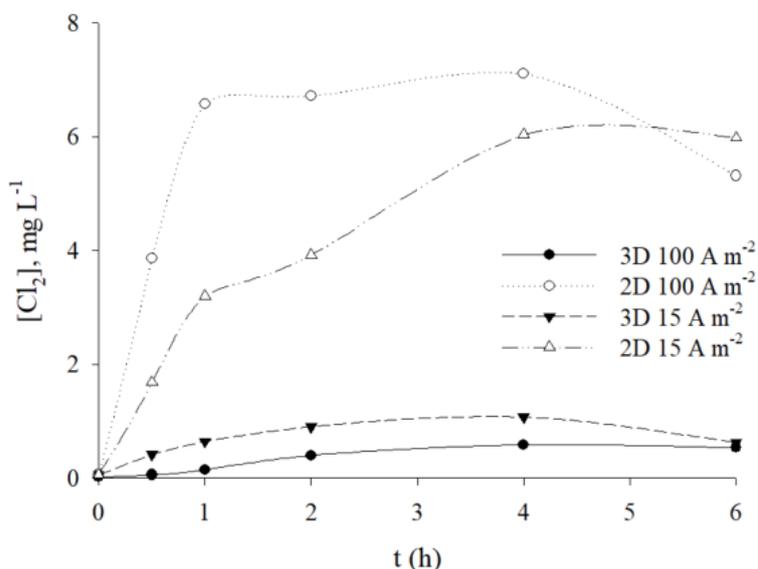

**Figure 5**. Concentrations of residual free chlorine as determined using the DPD method in 2D ELOX and 3D ELOX experiments, in the LCD and HCD experiments.

In the case of high oxidizing power anodes such as BDD, Cl$^-$ is further oxidized to ClO$_3^-$ and ClO$_4^-$ (Lan et al., 2017). In LCD experiments the (anode potential, $E_{AN}$=2.1 V/SHE), chlorate and perchlorate were measured in concentrations up to 0.19 and 0.13 mM in 2D ELOX, and 0.13 and 0.09 mM in 3D ELOX system, respectively (**Figure 4A, B**). Thus, addition of GAC packed bed did not have a significant impact on the concentrations of $ClO_3^-$ and $ClO_4^-$. Both species are formed by further anodic oxidation of chloride (eq. 5-9) (Bergmann et al., 2009):

$$2Cl^- \rightarrow Cl_2 + 2e^- \qquad \text{(eq. 5)}$$

$$Cl_2 + H_2O \rightarrow HClO + Cl^- + H^+ \qquad \text{(eq. 6)}$$

$$HClO \rightleftharpoons ClO^- + H^+ \qquad \text{(eq. 7)}$$

$$ClO^- + 2H_2O \rightarrow ClO_3^- + 4H^+ + 4e^- \qquad \text{(eq. 8)}$$

$$ClO_3^- + H_2O \rightarrow ClO_4^- + 2H^+ + 2e^- \qquad \text{(eq. 9)}$$

Previously, BDD anodic oxidation of secondary sewage effluent at 13 A m$^{-2}$ showed no $ClO_4^-$ formed during 3 h electrolysis, whereas $ClO_3^-$ was formed in concentrations of up to 0.15 mM (Cano et al., 2011). Nevertheless, it is difficult to compare the studies dealing with real wastewater treatment. The formation of perchlorate at low current densities may be inhibited by the presence of organics due to the competition between the organics and $ClO_3^\bullet$ for $OH^\bullet$ within a reaction zone (0.02—0.96 μm) adjacent to the BDD anode (eq. 10, 11) (Donaghue and Chaplin, 2013).

$$ClO_3^- \rightarrow ClO_3^\bullet + e^- \qquad \text{(eq. 10)}$$

$$ClO_3^\bullet + OH^\bullet \rightarrow HClO_4 \qquad \text{(eq. 11)}$$

Thus, application of low currents in BDD anodic oxidation of wastewater effluent did not avoid the generation of $ClO_3^-$ and $ClO_4^-$.

As expected, HCD experiments yielded higher chlorate and perchlorate concentrations, in 2D ELOX system, with $ClO_3^-$ reaching a concentration of 0.61 mM and $ClO_4^-$ of 1.18 mM at the end of the experiment (**Figure 4C**). In the 3D ELOX system (**Figure 4D**) the concentration of both was lower with concentration of $ClO_3^-$ reaching 0.30 mM after 6 h of electrolysis, whereas $ClO_4^-$ was 0.12 mM. Previous studies demonstrated efficient removal of $ClO_4^-$ by capacitive deionization (CDI) on activated carbon felt electrodes, and preferential adsorption of $ClO_4^-$ over $Cl^-$ (Xing et al., 2019). Enhanced electrosorption of perchlorate versus chlorate in our system can be explained by the higher dipole moment of $ClO_4^-$ and thus its higher polarizability (Howard et al., 1995; Siqueira et al., 2003). Thus, addition of activated carbon to electrochemical systems may be a suitable way to minimize the accumulation of toxic chlorinated byproducts due to their adsorption and electrosorption on GAC.

**Table 2** summarizes the measured concentrations of AOCl and AOI, whereas AOBr was below the detection limit of the method. The removal of AOI in 2D and 3D ELOX experiments at low applied current was similar, i.e., AOI concentration was decreased from 1,370 µg $L^{-1}$ to 901 and 813 µg $L^{-1}$, respectively. In the case of HCD, more intense formation of $OH^{\bullet}$ enhanced the degradation of iodinated organics and resulted in decreased concentrations of AOI, i.e., 439 and 307 µg $L^{-1}$ in 2D and 3D ELOX, respectively. Thus, addition of GAC packed bed led to an enhanced AOI removal only in the HCD experiments, likely due to enhanced electrosorption of typically high polarity iodinated organics. In terms of AOCl, the highest concentrations of AOCl measured in HCD experiments were, 576 µg $L^{-1}$ and 462 µg $L^{-1}$ for 2D and 3D ELOX, respectively. This is in accordance with the higher concentrations of residual free chlorine measured at higher current densities, and enhanced chloride oxidation, which thus lead to more intense electrochlorination of the organic matter (Schmalz et al., 2009; Xie et al., 2018; Yusufu Mohammed I., 2012). Nevertheless, addition of GAC packed bed lowered the AOCl concentration at both low and high current applied, i.e., from 498 µg $L^{-1}$ to 310 µg $L^{-1}$ in the LCD experiments, and from 576 µg $L^{-1}$ to 462 µg $L^{-1}$ in the HCD experiments.

**Table 2.** AOCl and AOI measured for the final sample (6 h) in ADS experiment, 2D ELOX and 3D ELOX experiments after 27 runs, for LCD (15 A m$^{-2}$) and HCD (100 A m$^{-2}$) experiments.

| | [AOCl], µg L$^{-1}$ | [AOI], µg L$^{-1}$ |
|---|---|---|
| **Secondary effluent** | 575 | 1370 |
| **ADS** | 442 | 743 |
| **2D ELOX, LCD** | 498 | 901 |
| **3D ELOX, LCD** | 310 | 813 |
| **2D ELOX, HCD** | 576 | 439 |
| **3D ELOX, HCD** | 462 | 308 |

## Conclusions

The performance of the 3D ELOX system with GAC packed bed was investigated at low and high current density (15 and 100 A m$^{-2}$, respectively). Synergy between electrosorption and electrochemical degradation was demonstrated when high current density applied, where the potential difference of the microconductor was large enough to induce efficient charge separation in GAC particles and electrochemical degradation of the (electro)sorbed contaminants. The highest synergy (up to 57%) was observed for DEET, the smallest molecule among the investigated contaminants and with a relatively high reactivity with HO$^{\bullet}$ (5.00×10$^9$). Surface characterization of GAC showed that at both low and high currents, porous structure of GAC undergoes changes, with an increased mesoporosity after exposing GAC to the electric field. Thus, the strength of the electric field impacts the structural properties of GAC and its ability of charge separation through polarization. Further, in-depth study should be conducted to evaluate the stability of different types of GAC and extent of changes in their porosity and surface area under bipolar polarization. The presence of GAC was demonstrated as beneficial for the reduction of halogenated organic by-products at both low and high currents applied. At 100 A m$^{-2}$, addition of GAC packed bed led to a significant decrease in the formed chlorate and perchlorate, from 0.61 to 0.30 mM of ClO$_3^-$ and from 1.18 to 0.12 mM of ClO$_4^-$ in 2D and 3D ELOX systems, respectively, likely due to their electrosorption onto the GAC granules. Thus, addition of relatively small amounts of activated carbon may be a feasible option for minimization of toxic and persistent inorganic chlorinated byproducts in electrochemical oxidation. Furthermore, in this study we intentionally excluded the development of a biofilm to investigate the synergy between electrosorption and electrochemical degradation. Yet, the performance of 3D ELOX system may be further enhanced by allowing the establishment of a biofilm and possibly additional contribution of biodegradation to the removal of organic pollutants,

**Acknowledgments**

This work has been funded by the ERC Starting Grant project ELECTRON4WATER (714177). ICRA researchers thank funding from CERCA program.

**Supplementary Material**

**Removal of persistent organic contaminants from wastewater using a hybrid electrochemical-granular activated carbon (GAC) system**

*Giannis-Florjan Norra[a,b], Jelena Radjenovic [a,c*]*

*[a]Catalan Institute for Water Research (ICRA), Emili Grahit 101, 17003 Girona, Spain*

*[b]University of Girona, Girona, Spain*

*[c]Catalan Institution for Research and Advanced Studies (ICREA), Passeig Lluís Companys 23, 08010 Barcelona, Spain*

*\* Corresponding author:*

*Jelena Radjenovic,*

*Catalan Institute for Water Research (ICRA), Emili Grahit 101, 17003 Girona, Spain*

Phone: + 34 972 18 33 80; Fax: +34 972 18 32 48; E-mail: jradjenovic@icra.cat

**Text S1 Chemical Analysis**

Target organic contaminants were analyzed with a 5500 QTRAP hybrid triple quadrupole-linear ion trap mass spectrometer with a turbo Ion Spray source (Applied Biosystems), coupled to a Waters Acquity Ultra-Performance$^{TM}$ liquid chromatograph (Milford). Iopromide (IPM), diatrizoate (DTR), N,N-diethyl-meta-toluamide (DEET) and carbamazepine (CBZ) were analyzed in electrospray (ESI) positive mode using an Acquity ultraperformance liquid chromatography (UPLC) HSS T3 column (2.1×50 mm, 1.8 μm, Waters) run at 30°C. The eluents employed were acetonitrile with 0.1% formic acid (eluent A), and milli-Q (LC-MS grade) water with 0.1% formic acid (eluent B) at a flow rate of 0.5 mL min$^{-1}$. The gradient was started at 2% of eluent A that was increased to 20% A by 3 min, further increased to 50% A by 6 min and further increased to 95% A by 7 min. It was kept constant for 2.5 min, before returning to the initial condition of 2% A by 9.5 min. The total run time was 11 min. The target organic contaminants were analyzed in a multiple reaction monitoring (MRM). The settings for the compound-dependent parameters of each transition are summarized in Table S3.

**Table S1** Chemical structures and physico-chemical properties of target contaminants; molecular weight (MW), pKa, octanol-water distribution coefficient calculated based on chemical structure at pH 7.4 (CX LogD), polar surface area (PSA) and polarizability, i.e., ability to form instantaneous dipoles. Calculated CX logD values were collected by ChEMBL database. Polar surface areas and polarizability were collected from Chemspider.com database.

| Organic compound (MW, g mol⁻¹) | Chemical structure | pKa | PSA, Å² | Polarizability, Å³ | logD |
|---|---|---|---|---|---|
| Iopromide (791.1) |  | 9.9[1] | 169 | 56.4 | -0.44 |
| Carbamazepine (236.3) |  | 13.9[2] | 47 | 28 | 2.77 |
| Diatrizoate (613.9) |  | 3.4[3] | 96 | 39.8 | -0.63 |
| DEET (191.3) |  | 0.67[4] | 20 | 23.3 | 2.50 |

**Table S2** Characterization of secondary treated sewage effluent.

| Parameter | pH | Electric conductivity ($\mu$S cm$^{-1}$) | COD (mg L$^{-1}$) | TOC (mg L$^{-1}$) | Chloride (mg L$^{-1}$) |
|---|---|---|---|---|---|
| **Value** | 8.0±0.04 | 1154±18 | 55.7±9.6 | 13.6±0.5 | 177.5±0.5 |

**Table S3** The optimized compound-dependent MS parameters: declustering potential (DP), collision energy (CE) and cell exit potential (CXP)) for each compound and each transition of the negative and positive mode.

| Organic compound | Q1 Mass (Da) | Q3 Mass (Da) | DP | CE | CXP |
|---|---|---|---|---|---|
| **Iopromide** | 791.72 | 572.9 | 156 | 35 | 20 |
| | 791.72 | 300 | 156 | 83 | 10 |
| **Carbamazepine** | 237.01 | 194.1 | 156 | 47 | 10 |
| | 237.01 | 193 | 156 | 47 | 10 |
| **DEET** | 193.05 | 120 | 176 | 23 | 14 |
| | 193.05 | 91.6 | 176 | 39 | 12 |
| **Diatrizoate** | 614.9 | 361 | 80 | 30 | 10 |
| | 614.9 | 233.1 | 85 | 33 | 10 |

**Table S4** Apparent first-order removal rate constants ($h^{-1}$) of the target contaminants at 15 A $m^{-2}$ (LCD experiment) and 100 A $m^{-2}$ (HCD experiment) in 2D ELOX and 3D ELOX system in different runs.

| | 2D ELOX | ADS | RUN1 | RUN 6 | RUN 11 | R12 ADS | RUN 27 | R28 ADS |
|---|---|---|---|---|---|---|---|---|
| **IPM** | | | | | | | | |
| **HCD** | 0.17±0.02 | 0.08±0.01 | 0.32±0.08 | 0.37±0.054 | 0.45±0.03 | 0.10±0.01 | 0.46±0.12 | 0.13±0.002 |
| **LCD** | 0.10±0.07 | | 0.20±0.06 | 0.19±0.03 | 0.12±0.07 | 0.029±0.002 | 0.16±0.05 | 0.023±0.002 |
| **CPZ** | | | | | | | | |
| **HCD** | 0.36±0.10 | 0.33±0.08 | 0.81±0.15 | 0.89±0.01 | 0.96±0.01 | 0.41±0.04 | 1.05±0.02 | 0.43±0.03 |
| **LCD** | 0.12±0.05 | | 0.70±0.06 | 0.59±0.03 | 0.36±0.07 | 0.17±0.01 | 0.33±0.05 | 0.08±0.01 |
| **DTR** | | | | | | | | |
| **HCD** | 0.22±0.07 | 0.02±0.01 | 0.25±0.04 | 0.31±0.02 | 0.32±0.04 | 0.027±0.014 | 0.34±0.07 | 0.032±0.009 |
| **LCD** | 0.12±0.09 | | 0.13±0.02 | 0.17±0.02 | 0.12±0.01 | 0.017±0.004 | 0.16±0.05 | 0.014±0.007 |
| **DEET** | | | | | | | | |
| **HCD** | 0.17±0.02 | 0.19±0.07 | 0.60±0.19 | 0.68±0.09 | 0.83±0.10 | 0.20±0.08 | 0.84±0.16 | 0.30±0.03 |
| **LCD** | 0.12±0.07 | | 0.46±0.30 | 0.51±0.01 | 0.32±0.05 | 0.16±0.02 | 0.33±0.08 | 0.15±0.03 |

**Table S5** Synergies between adsorption and electrochemical oxidation for the several runs in the HCD experiments.

| | RUN1 | RUN 6 | RUN 11 | RUN 27 |
|---|---|---|---|---|
| **IPM** | 22,2% | 32,3% | 44,4% | 45,5% |
| **CBZ** | 14,6% | 22,2% | 28,1% | 34,0% |
| **DTR** | 4,4% | 22,7% | 24,4% | 29,8% |
| **DEET** | 40,1% | 47,4% | 57,2% | 57,4% |

**Table S6** Bimolecular rate constants (k, M$^{-1}$ s$^{-1}$) for oxidation of target organic contaminants with homogeneously generated OH$^•$, O$_3$ and Cl$_2$.

| Contaminant | k$_{•OH}$, M$^{-1}$ s$^{-1}$ | k$_{O3}$, M$^{-1}$ s$^{-1}$ | k$_{Cl2}$, M$^{-1}$ s$^{-1}$ |
|---|---|---|---|
| **Iopromide** | $3.30 \times 10^{9[5]}$ | 93.1[6] | 0[9] |
| **Carbamazepine** | $8.80 \times 10^{9[5]}$ | $3 \times 10^{5[7]}$ | <0.1[10] |
| **Diatrizoate** | $6.3 \times 10^{8[5]}$ | 48.6[6] | 0[9] |
| **DEET** | $5.00 \times 10^{9[5]}$ | 0.1[8] | $1.3 \times 10^{-3}$ [11] |

**Table S7** Electric energy per order ($E_{EO}$, kWh m$^{-3}$) during the several runs

| $E_{EO}$ (kWh m$^{-3}$) | 2D ELOX | RUN 1 | RUN 6 | RUN 11 | RUN 27 |
|---|---|---|---|---|---|
| **IPM** | - | - | 41.9 | 46.2 | 42.8 |
| **CBZ** | 42.3 | 14.4 | 17.5 | 23.1 | 16.4 |
| **DTZ** | - | - | - | - | - |
| **DEET** | - | 20.6 | 22.4 | - | - |

The values with (-) are missing because one order of removal was not achieved

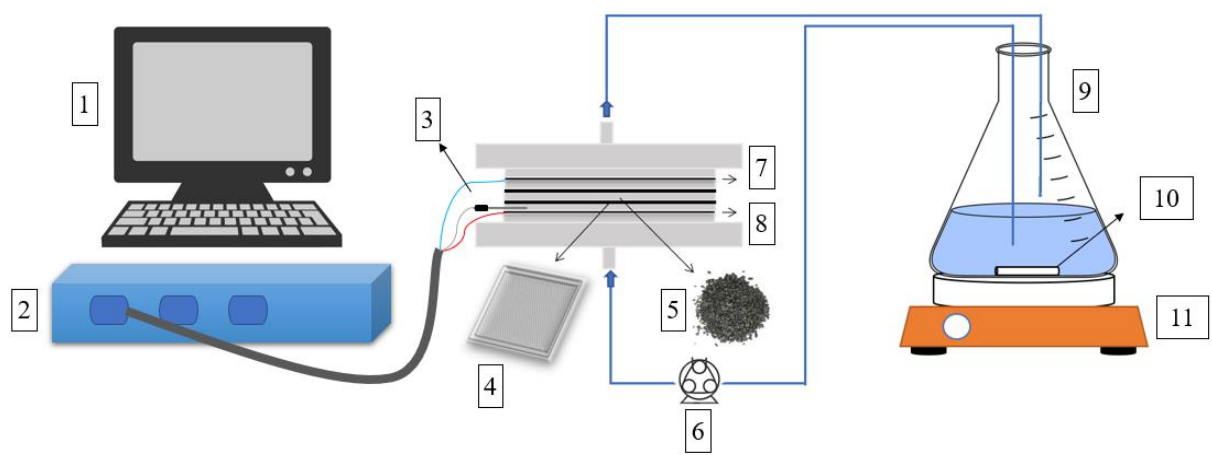

**Figure S1**. Scheme of the experimental set-up used in adsorption (ADS), two-dimensional electrooxidation (2D ELOX) and three-dimensional electrooxidation (3D ELOX) experiments. (1) computer for control of current and data acquisition, (2) potentiostat, (3) reference electrode, (4) inert mesh holder, (5) granules of the granular activated carbon (GAC) packed bed, (6) peristaltic pump, (7) stainless steel cathode, (8) boron-doped diamond (BDD) anode, (9) secondary effluent reservoir, (10) magnet, (11) magnetic stirrer.

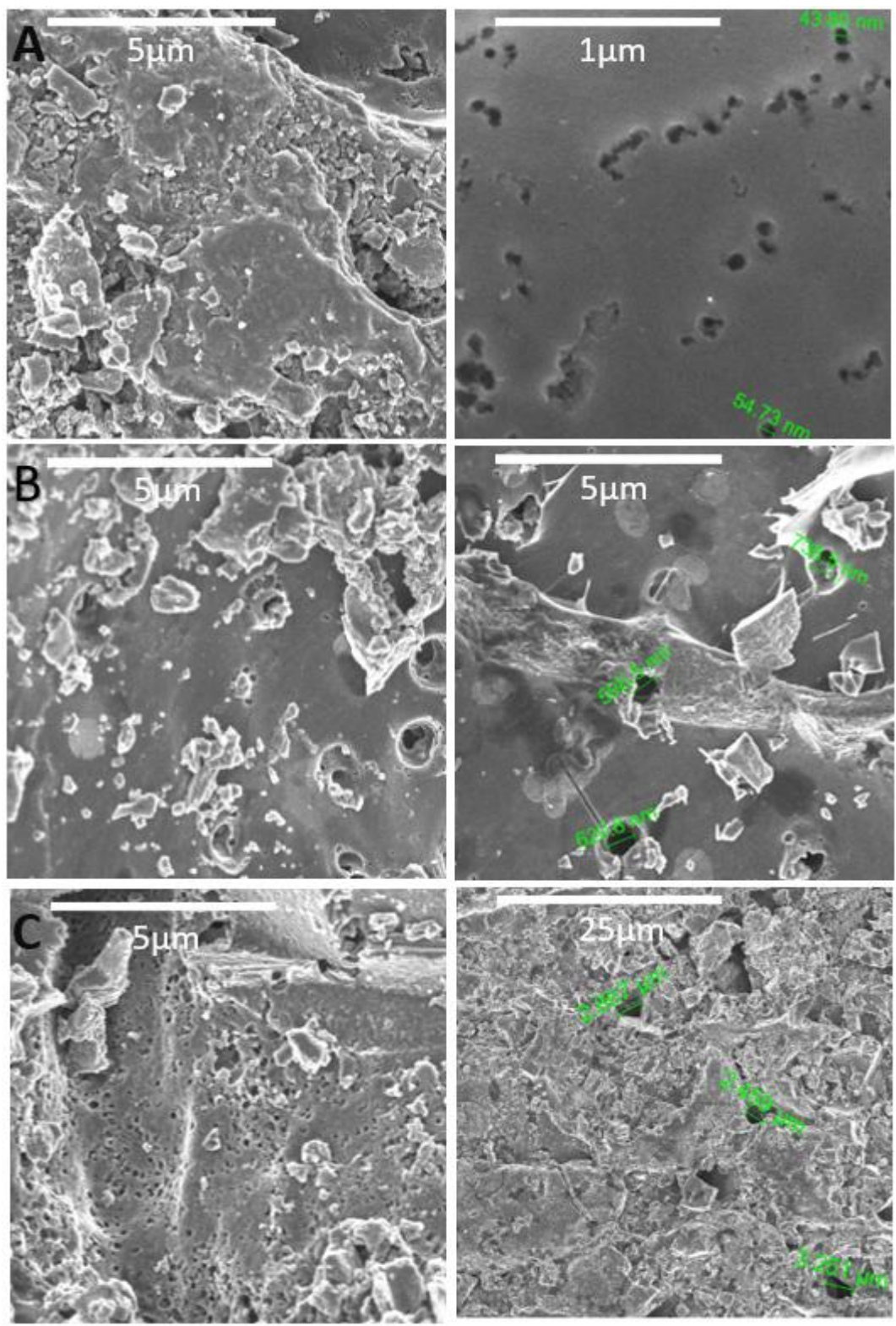

**Figure S2** SEM images: **A)** initial, saturated GAC granules before electrochemical polarization, **B)** GAC granules after the LCD treatment, and **C)** GAC granules after HCD treatment